\documentclass{optica-article}

\journal{opticajournal} 

\articletype{Research Article}

\usepackage{lineno}

\begin{document}

\title{Stochastic logic in biased coupled photonic probabilistic bits}

\author{Michael Horodynski,\authormark{1,*} Charles Roques-Carmes,\authormark{2,3} Yannick~Salamin,\authormark{1,3} Seou Choi,\authormark{3} Jamison Sloan,\authormark{3} Di Luo,\authormark{1,4,5} and Marin Solja\v{c}i\'{c}\authormark{1,3}}

\address{\authormark{1}Department of Physics, Massachusetts Institute of Technology, Cambridge, MA 02139, USA\\
\authormark{2}E. L. Ginzton Laboratory, Stanford University, Stanford, CA 94305, USA\\
\authormark{3}Research Laboratory of Electronics, Massachusetts Institute of Technology, Cambridge, MA 02139, USA\\
\authormark{4}The NSF AI Institute for Artificial Intelligence and Fundamental Interactions\\
\authormark{5}Department of Physics, Harvard University, Cambridge, MA 02138, USA}
\email{\authormark{*}mhorodyn@mit.edu}


\begin{abstract*}
Optical computing often employs tailor-made hardware to implement specific algorithms, trading generality for improved performance in key aspects like speed and power efficiency. An important computing approach that is still missing its corresponding optical hardware is probabilistic computing, used e.g. for solving difficult combinatorial optimization problems. In this study, we propose an experimentally viable photonic approach to solve arbitrary probabilistic computing problems. Our method relies on the insight that coherent Ising machines composed of coupled and biased optical parametric oscillators can emulate stochastic logic. We demonstrate the feasibility of our approach by using numerical simulations equivalent to the full density matrix formulation of coupled optical parametric oscillators.
\end{abstract*}


For the past 10 years, there has been a strong renewed interest in optical computing. The reasons are three-fold: (1) the exploration of alternative hardware approaches for essential applications like neural networks; (2) a mismatch between future computing resources demand and hardware progress due to an impending slowdown of Moore's law; and (3) improvement of optical hardware \cite{mcmahon_physics_2023}. A paradigmatic example is that of a fully connected neural network implemented in multi-layer networks of Mach-Zehnder interferometers \cite{shen_deep_2017}. Several important classes of computing frameworks are, however, still missing their matching optical computing hardware, and one of them is probabilistic computing \cite{kirkpatrick_optimization_1983,geman_stochastic_1984}. In probabilistic computing ($p$-computing), probabilistic bits ($p$-bits) replace the deterministic bits of conventional computing, while still sharing the use of basic logic gates that can be assembled to construct more complex circuits \cite{camsari_stochastic_2017,borders_integer_2019}. The framework of stochastic logic allows for the creation of ``invertible'' logic circuits (circuits that cannot only produce an output from some input but can also take the output to go back to the inputs) \cite{camsari_stochastic_2017}. This makes $p$-computing especially suitable for solving complex optimization problems \cite{aadit_massively_2022,reifenstein_coherent_2023,smithson_efficient_2019,zhang_solving_2021} and for simulating physical systems, like the Ising Hamiltonian, which represents a collection of spins with arbitrary (linear) inter-spin coupling \cite{mcmahon_fully_2016,vadlamani_physics_2020,roques-carmes_heuristic_2020}. Furthermore the use of $p$-computing has been demonstrated for a wide array of different applications, including Bayesian inference \cite{harabi_memristor-based_2022}, fast restricted Boltzmann machines, classical annealing, \cite{camsari_p-bits_2019}, quantum Monte Carlo \cite{chowdhury_accelerated_2023}, machine learning quantum many-body systems \cite{chowdhury_full-stack_2023} and generative neural networks \cite{choi_photonic_2024}.

Concurrently, there has been a significant interest in leveraging optics and other physical platforms for the determination of ground states in Ising Hamiltonians, a technique commonly referred to as coherent Ising machines \cite{mohseni_ising_2022}. A pioneering study proposed the utilization of injection-locked laser systems to map Ising models~\cite{utsunomiya_mapping_2011}. Subsequently, the proposition and experimental realization of networks of coupled optical parametric oscillators (OPOs) emerged \cite{wang_coherent_2013,maruo_truncated_2016,yamamoto_coherent_2017}, with implementations reported in Refs.~\cite{marandi_network_2014,mcmahon_fully_2016,inagaki_coherent_2016,honjo_100000-spin_2021,okawachi_demonstration_2020}. Additionally, alternative approaches in optics and photonics have been considered to solve Ising problems \cite{wu_optical_2014,vazquez_optical_2018,pierangeli_large-scale_2019,jacucci_tunable_2022}.
The goal of these studies is to realize OPO networks described by the general Ising Hamiltonian of the form
\begin{equation}
    E = -\tfrac{1}{2}\sum_{i,j}\sigma_i J_{ij} \sigma_j - \sum_i h_i \sigma_i,
\end{equation}
where $J$ is the coupling matrix between different spins, represented by $\Vec{\sigma}$ and $\Vec{h}$ is the magnetization (Zeeman) vector. It is important to note, that the aforementioned studies only considered Ising Hamiltonians without a local magnetic field ($\Vec{h}=0$). Coherent Ising machines are a prime candidate for implementing $p$-computing in the optical domain since a network of $p$-bits can be mapped to an Ising Hamiltonian \cite{camsari_stochastic_2017}. However, to fully unlock the potential of $p$-computing in optics, coherent Ising machines that can only consider the coupling $J$ are insufficient. This can be readily understood by considering even a simple logic gate, such as the AND gate, which, when implemented with stochastic logic, necessitates a non-zero Zeeman term. It is therefore important to develop strategies that incorporate a Zeeman term, with realizations as a feedback signal \cite{takesue_simulating_2020} or as an artificial Zeeman term \cite{inui_control_2022}. A recent experiment~\cite{roques-carmes_biasing_2023} showed that injecting a vacuum-level bias field in an OPO cavity could be used to control its steady-state distribution, akin to the action of a magnetic field on a single spin. This provides a new strategy to include a Zeeman term with external coherent fields.

\begin{figure}[t!]
    \centering
    \includegraphics[width=\textwidth]{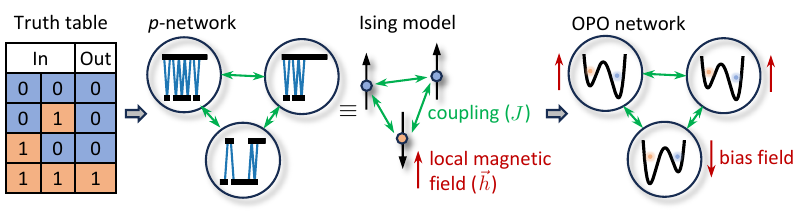}
    \caption{Illustration of the three-step conversion process. We take the truth table from a (basic) digital logic gate and map it to a network of $p$-bits. This amounts to describing the gate as an Ising model with a coupling term ($J$) and a local magnetic field ($\vec{h}$). We then find the ground state of the Ising model by looking at the steady state of a network of (i.e. coupled) optical parametric oscillators (OPOs) that are biased by an injected field.}
    \label{fig:1}
\end{figure}

Here, we propose an optical platform for $p$-computing that consists of networks of coupled and biased OPOs to emulate arbitrary stochastic logic circuits. Our method relies on a three-step process that maps a stochastic logic problem to the dynamics of a biased OPO network. The process starts with (1) a logic truth table listing all possible outcomes of the logic gate or task; (2) converting that truth table into an equivalent Ising model (including a non-zero Zeeman term); and (3) finding an equivalent OPO network that realizes the corresponding Ising Hamiltonian (see Fig.~\ref{fig:1}). The motivation for our proposal of OPOs as optical hardware for $p$-computing is twofold: On one hand, the nature of $p$-bits to randomly fluctuate between either 0 or 1 is reminiscent of the OPO (a nonlinear, stochastic, and dissipative system) feature to choose its steady state randomly between two possible phases (often called $\pm$ 1). On the other hand, networks of OPOs have been shown to map their steady-state configurations to the ground states of Ising Hamiltonians, which has garnered significant interest in the optical computing community~\cite{wang_coherent_2013}. The main innovation of our work is to realize the Zeeman term by biasing each node of the OPO network, which is a key ingredient in realizing all-purpose stochastic logic circuits.

We now demonstrate why such a bias field indeed allows for the determination of the Ising ground state. Physically the bias field is a coherent field injected into the OPO cavity at the signal frequency. It influences the OPO's steady state by displacing the initial vacuum state from its mean value of zero (which gets amplified to choosing one of two possible steady states with equal probability) to non-zero mean value, which then gets amplified to a tunable (by strength of the bias field) binomial probability distribution. Our starting point is the set of stochastic differential equations describing the OPO dynamics, which are rigorously equivalent to the density matrix formulation of degenerate parametric oscillation in the presence of a bias field \cite{haribara_computational_2016}:
\begin{align}
    \mathrm{d}c_i &= \left(p - 1 - c_i^2 - s_i^2 \right)c_i\mathrm{d}t + \varepsilon \left( \sum_{j}g_{ij}c_j + b_i\right)\mathrm{d}t + \frac{1}{A_s}\sqrt{\tfrac{1}{2} + c_i^2 + s_i^2}\mathrm{d}W_c, \label{eq:2}\\
    \mathrm{d}s_i &= \left(- p - 1 - c_i^2 - s_i^2 \right)s_i\mathrm{d}t + \varepsilon \sum_{j}g_{ij}s_j \mathrm{d}t + \frac{1}{A_s}\sqrt{\tfrac{1}{2} + c_i^2 + s_i^2}\mathrm{d}W_s \label{eq:3}. 
\end{align}
Here, $c_i$ is the in-phase and $s_i$ the quadrature component of the $i$th OPO, $p$ the pump, the saturation amplitude $A_s$ is the field amplitude at a pump rate $p=2$, $\mathrm{d}W_c$ and $\mathrm{d}W_s$ are two independent Gaussian noise processes and $\varepsilon$ controls the overall strength of coupling $g_{ij}$ and bias $b_i$. Note that we have chosen the bias to be in phase with $c_i$. In the following, we demonstrate how the OPO network finds the ground state of the Ising model including a Zeeman term. To this end, we neglect the noise terms, since we are interested in the mean fields at steady state. As a first step, we note that also in the multi-OPO case, the quadrature component is zero $s_i=0 \; \forall\; i$, just like it is for a single OPO above threshold pumping \cite{wang_coherent_2013}. The second step is (following Ref.~\cite{wang_coherent_2013}) to look at the overall photon decay rate at the steady state ($\mathrm{d}c_i=\mathrm{d}s_i=0$), defined as (note that we have already used the fact here that $s_i=0$):
\begin{equation}
    \Gamma = \sum_i(p-c_i^2).
\end{equation}
We can calculate $\Gamma$ by expanding $c_i$ into orders of $\varepsilon$, since we assume a small magnitude for each $g_{ij}$ and $b_i$
\begin{align}
    c_i & = c_i^{(0)} + \varepsilon c_i^{(1)} + \dots,\; \mathrm{with}\\
    c_i^{(0)} & = \sqrt{p-1} \sigma_i \quad \mathrm{and} \quad
    c_i^{(1)} = \frac{1}{2(p-1)}\left( \sqrt{p-1} \sum_j g_{ij} \sigma_j  + b_i \right), \label{eq:6}
\end{align}
where the spin configuration is given by $\sigma_i = \pm 1$. We arrive at the components for the expansion of the in-phase component $c_i$ [Eq.~\eqref{eq:6}] by iteratively solving Eq.~\eqref{eq:2}. From this, the overall photon decay rate then readily follows as
\begin{align}
    \Gamma & = N - \varepsilon \left( \sum_{i,j} g_{ij}\sigma_i\sigma_j + \frac{1}{\sqrt{p-1}}\sum_i b_i \sigma_i \right), 
\end{align}
which is largest if the OPO network is in the ground state of the Ising Hamiltonian, assuming that we replaced $g_{ij} \rightarrow \tfrac{1}{2} J_{ij}$ and $b_i \rightarrow \sqrt{p-1} h_i$.

However, simply solving the system described by Eqs.~\eqref{eq:2} and \eqref{eq:3} is not sufficient for finding the ground state due to heterogeneity of the amplitudes $c_i$. This is due to improper mapping of the objective functions by the loss landscape \cite{leleu_destabilization_2019}. Following Ref.~\cite{chen_cim-optimizer_2022}, we fix this by adding an additional dynamic error field $e_i$, resulting in the following set of equations of motion:
\begin{align}
    \mathrm{d}c_i &= \left[ \left(p - 1 - c_i^2 - s_i^2 \right)c_i + \varepsilon e_i \left( \sum_{j}J_{ij}c_j + F_h h_i\right)\right]\mathrm{d}t + \frac{1}{A_s}\sqrt{\tfrac{1}{2} + c_i^2 + s_i^2}\mathrm{d}W_c, \label{eq:8} \\
    \mathrm{d}s_i &= \left[ \left(- p - 1 - c_i^2 - s_i^2 \right)s_i + \varepsilon e_i \sum_{j}J_{ij}s_j \right] \mathrm{d}t + \frac{1}{A_s}\sqrt{\tfrac{1}{2} + c_i^2 + s_i^2}\mathrm{d}W_s, \label{eq:9} \\
    \frac{\mathrm{d}e_i}{\mathrm{d}t} & = -\beta \left( c_i^2 - a \right)e_i. \label{eq:10}
\end{align}
Here, $e_i$ is an auxiliary field that helps us homogenize the amplitudes (with target $a$) from the different OPOs to reduce the number of stable local minima, $\beta$ controls the rate of change in $e_i$ and $F_h$ is the bias amplitude.

\begin{figure}[t!]
    \centering
    \includegraphics[width=\textwidth]{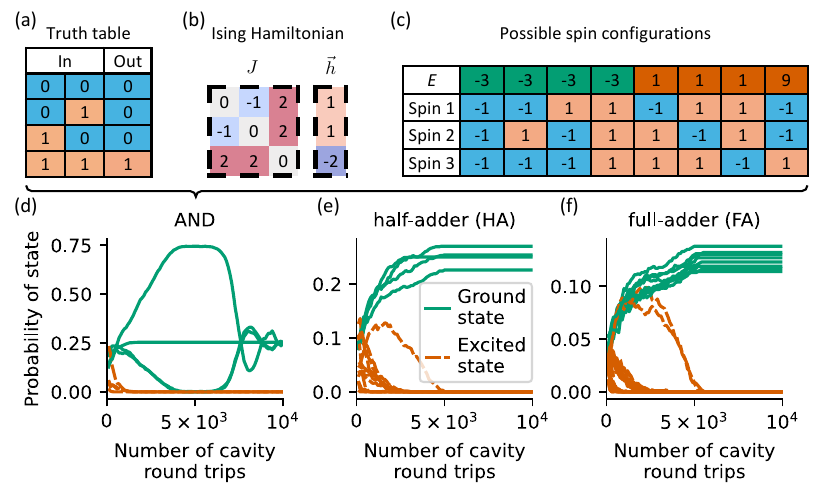}
    \caption{Fundamental p-gates implemented with a biased coupled OPO network. (a) Truth table of the AND-gate. (b) Corresponding Ising Hamiltonian with three spins. (c) All possible spin configurations of the Hamiltonian depicted in (b) with their corresponding energy $E$. The ground states reproduce the correct logic from the truth table. (d)-(e) We plot the probability for each possible spin configuration of the three (AND), four (HA), and five (FA) spins as a function of the number of cavity round-trips. Depicted is the probability that the OPO network is currently in a ground state of the corresponding Ising Hamiltonian (green), and in an excited state (dark orange, dashed).}
    \label{fig:2}
\end{figure}

To show that the aforementioned system of coupled OPOs is indeed a viable tool to implement stochastic gates, we now consider three representative fundamental gates: The first is the AND gate, which allows for the multiplication of two 1-bit numbers. The second is the half-adder that can take two 1-bit numbers and outputs the resulting sum and a carry-bit. The third is the full-adder that can take two 1-bit numbers and a carry-bit and outputs the resulting sum plus a carry-bit. The truth table associated with each of the three gates are the possible spin configurations of the corresponding Ising Hamiltonian's ground state with three, four, and five spins respectively \cite{whitfield_ground-state_2012,pervaiz_hardware_2017,onizawa_design_2021}. In each case, both the bits on the input and output sides of the binary logic truth table are represented by OPOs in the same network. In Fig.~\ref{fig:2} we show a detailed description of how the AND gate can be implemented in stochastic logic framework and the time evolution of the occupation probability for a certain spin configuration for the three gates under consideration. We observe that after some time the spin configuration in each OPO network represents a ground state of the Ising Hamiltonian with approximately equal probability, while simultaneously no OPO network is in an excited state. We note here that an OPO network can thus (in principle) perform any arbitrary computing task that can be represented by a logic circuit. In addition to the possibility of general computing, there is also the opportunity to implement invertible logic, i.e. the ability to go through logic circuits from output to input \cite{camsari_stochastic_2017}. This insight now allows us to go one step further and construct more complex stochastic logic circuits, allowing us to solve difficult combinatorial optimization problems.

\begin{figure}[t!]
    \centering
    \includegraphics[width=\textwidth]{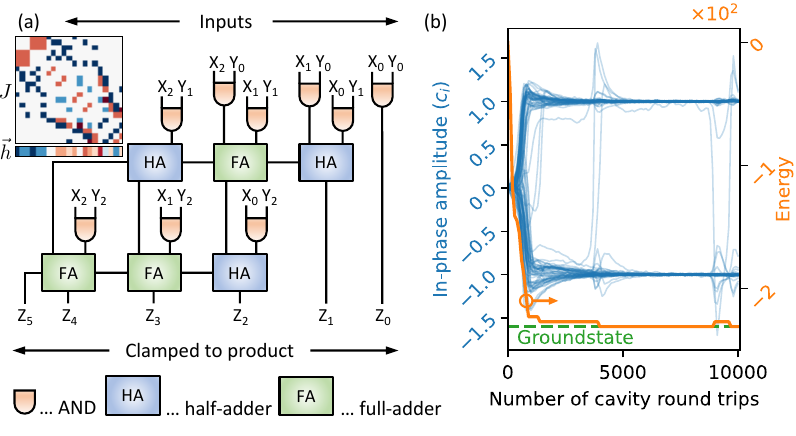}
    \caption{Solving the semiprime factorization problem with a coupled and biased network of OPOs. (a) Digital logic circuit that multiplies two 3-bit numbers ($X$ and $Y$ in their binary representations, respectively), resulting in $Z$ (binary representation). The lines interconnecting the gates are represented by auxiliary spins in the Ising Hamiltonian. The inset shows the corresponding Ising Hamiltonian with coupling $J$ at the top and Zeeman term $\vec{h}$ at the bottom. (b) Time evolution of the in-phase component ($c_i$) of coupled and biased OPOs with $J$ and $\vec{h}$ originating from the 6-bit multiplier circuit. The steady state of this configuration represents the ground state of the Ising Hamiltonian, solving the factorization problem.}
    \label{fig:3}
\end{figure}

As a first example, we choose a particular problem that has been established as an appropriate testing ground for probabilistic computing: semiprime factorization. Given a number, we aim to find the two prime numbers that, when multiplied, are equal to this number. This problem is of great interest since no polynomial time algorithms on classical computers have been discovered so far, making it for example the basis of several cryptographic systems~\cite{shor_polynomial-time_1997}. We show in Fig.~\ref{fig:3}a a possible implementation of a logic circuit that takes 2 numbers and outputs their product. Conventionally this circuit works one way from top to bottom. However, the framework of stochastic logic allows us to clamp the output to the number we want to factorize in order to find the factors as the ground state of the associated Ising Hamiltonian (shown as an inset in Fig.~\ref{fig:3}a). This Hamiltonian can be automatically generated by stitching together the Hamiltonians of the basic gates we demonstrated in Fig.~\ref{fig:2}. We want to emphasize that this procedure scales linearly with the number of bits required to represent the integers involved. Furthermore, we point out that we gave an example of a general constructive method to map most computing problems (as long as they can be cast as a logic circuit) into a network of OPOs.

To demonstrate that our network of coupled and biased OPOs can indeed find the ground state of the Hamiltonian depicted in Fig.~\ref{fig:3}a we now factorize $2491 = 47 \times 53$. Since both multiplicands are 6-bit numbers, altogether 12 OPOs (in conjunction with 84 OPOs representing the interconnects in the circuit) allow us to read out the solution. In Fig.~\ref{fig:3}b we show the trajectory of the in-phase components ($c$) of each OPO in the network and observe that the spin configuration given by $\sigma_i = c_i/|c_i|$ settles into the ground state of the Ising Hamiltonian. The solution to the semiprime factorization problem is then encoded by the spin configuration in a binary representation. As a rule of thumb, we observe that the number of cavity round trips required to find the ground state stays independent of the number of OPOs involved. For example, with a $80$MHz repetition rate laser (as used in Ref.~\cite{roques-carmes_biasing_2023}) one loop around the OPO cavity is approximately $3.75$m long. It then takes $0.125$ms to perform 10000 cavity round trips. 

\begin{figure}[t!]
    \centering
    \includegraphics[width=\textwidth]{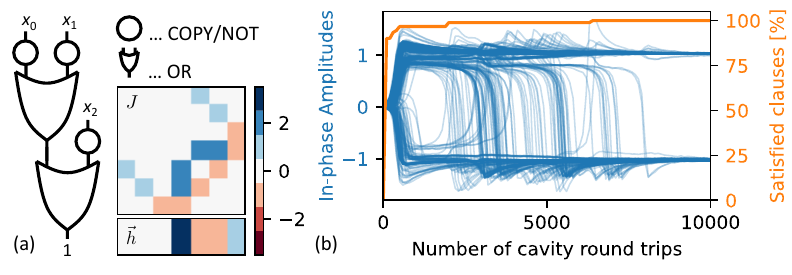}
    \caption{Solving the Boolean satisfiability ($\text{SAT}$) problem. (a) The digital logic circuit of one clause in the $3\text{SAT}$ problem is shown with the corresponding Ising Hamiltonian. (b) Temporal evolution towards the steady state of the in-phase component ($c$, blue lines) of 384 coupled and biased OPOs with $J$ and $\vec{h}$ originating from the $3\text{SAT}$ problem with 20 variables and 91 clauses. From the steady state, we can read off one solution to the $3\text{SAT}$ problem. The orange line plots the evolution of the percentage of clauses satisfied.}
    \label{fig:4}
\end{figure}

We show our approach's flexibility by also solving a different NP-complete problem with a coupled and biased OPO network: boolean satisfiability ($\text{SAT}$). More concretely, we consider the $3\text{SAT}$ problem, in which we want to find configurations ($\vec{x}$) that satisfy a conjunction of clauses (here, 91 clauses) of exactly three terms, which may or may not be negated. For instance, one such clause may take the form $x_i \lor x_j \lor \overline{x_k}$, where $\lor$ denotes the logical ``or'' and $\overline{x_i}$ represents a negation. A simple example problem with 2 clauses and 4 variables is $\left(x_1 \lor \overline{x_2} \lor x_4 \right) \land \left( x_2 \lor \overline{x_3} \lor \overline{x_4} \right)$, with $\land$ denoting the logical ``and''. Fig.~\ref{fig:4}a depicts the logic circuit for one clause of the 3SAT problem, with the output clamped to $1$ so that the invertible logic property of $p$-computing allows for finding the state of $\vec{x}$ that satisfies this clause. Between the variables $x_i$ and the OR gates, we put either a COPY or NOT-gate, depending on whether the variable is or is not negated in the clause. We note here that $p$-computing is especially suited to solve the $3\text{SAT}$, since we only have linear overhead to represent the problem as an Ising Hamiltonian, stemming from $N_\mathrm{spins} = N_\mathrm{variables} + 4 N_\mathrm{clauses}$. Here, $N_\mathrm{spins}$ (resp., $N_\mathrm{variables}$, $N_\mathrm{clauses}$) is the number of spins (resp., variables, clauses).

Fig.~\ref{fig:4}b shows the in-phase component's ($c$) time evolution in an OPO network configured to solve the ‘uf20-01.cnf’ $\text{SAT}$ problem (a common benchmark example) with $N_\mathrm{variables} = 20$ and $N_\mathrm{clauses} = 91$ \cite{hoos_satlib_2000}. Plotted is also the percentage of satisfied clauses as a function of time and observe that a configuration of variables that satisfies all clauses is found before the steady state is reached. This can be attributed to the fact that the OPOs representing the variables are already in the correct state while the states of OPOs representing the interconnects in the circuit are still changing their state. We note that the parameters in the full set of equations describing the system of coupled and biased OPOs, including amplitude heterogeneity correction [Eqs.~\eqref{eq:8}-\eqref{eq:10}], require careful tuning for the system to find the ground state of the Ising Hamiltonian. An intuitive guideline for selecting these parameters is to ensure that the pump strength $p$ and the overall coupling strength $\varepsilon$ are balanced. This balance prevents the system from being dominated solely by the pump, where the coupling bias field becomes negligible, or by the coupling and biasing alone. Additionally, the bias amplitude $F_h$ should be set so that the coupling $J$ and the magnetic field $\vec{h}$ are balanced. The rate of change $\beta$ in the auxiliary field $\Vec{e}$ also needs to be chosen carefully. This ensures that the amplitude heterogeneity correction neither overtakes the evolution to the steady state nor fails to guide the system effectively to the ground state of the Ising Hamiltonian.

To conclude, we propose an approach that allows for the implementation of arbitrary stochastic logic circuits in a network of coupled and biased OPOs, working closely along the lines of experimentally demonstrated technology. We successfully tested the approach numerically for a representative set of basic logic circuits and ubiquitous problems in combinatorial optimization. Our study paves the way for an optical implementation of $p$-computing for the potentially high-speed solution of ubiquitous combinatorial optimization problems.

For an experimental realization of our approach, we envision a set-up similar to the one used in the first successful experiment on applying a bias field to an OPO \cite{roques-carmes_biasing_2023}. Combining it with a single fiber-ring cavity then provides the missing ingredient: coupling between the (time-multiplexed) OPOs through feedback and measurement \cite{mcmahon_fully_2016}. Ultimately, however, we anticipate the coupling between different OPOs to be performed fully optically by e.g. linear programmable nanophotonic processors for great speed and energy efficiency \cite{harris_linear_2018}. Generally, $p$-computing would also profit from massively parallel, ultrafast, and tunable random bit generation using a physical source for the randomness \cite{kim_massively_2021}.

\begin{backmatter}
\bmsection{Funding}
M.~H. is funded by the Austrian Science Fund (FWF) through grant J 4729-N. C.~R.-C. is supported by a Stanford Science Fellowship. S.~C. acknowledges support from Korea Foundation for Advanced Studies Overseas PhD Scholarship. D.~L., and M.~S. acknowledge support from the National Science Foundation under Cooperative Agreement PHY-2019786 (The NSF AI Institute for Artificial Intelligence and Fundamental Interactions, \url{http://iaifi.org/}). This work is also supported in part by the U. S. Army Research Office through the Institute for Soldier Nanotechnologies at MIT, under Collaborative Agreement Number W911NF-23-2-0121.

\bmsection{Acknowledgments}
The authors acknowledge the MIT SuperCloud and Lincoln Laboratory Supercomputing Center for providing high-performance computing resources that have contributed to the research results reported within this paper.

\end{backmatter}



\end{document}